\documentclass[twocolumn]{aa}
\usepackage{subfig}
\usepackage{float}
\usepackage{graphicx}
\usepackage{txfonts}
\usepackage{natbib}
\usepackage{times}
\begin{document}
   \title{Bar pattern speed evolution over the last 7 Gyr}


   \author{I. P\'erez, 
          \inst{1,2}
          J.A.L. Aguerri\inst{3,4}
          \and
          J. M\'endez-Abreu\inst{3,4}
          }
    \institute{Dpto. de F\'isica Te\'orica y del Cosmos, University of Granada, Facultad de Ciencias (Edificio Mecenas), 18071-Granada, Spain\\
              \email{isa@ugr.es}
         \and
             Instituto Carlos I de F\'isica Te\'orica y Computaci\'on
         \and
              Instituto de Astrof\'{\i}sica de Canarias, C/ V\'{\i}a L\'actea s/n, 38200 La Laguna, Spain\\
              \email{jalfonso@iac.es, jairo@iac.es}
                         \and
              Departamento de Astrof\'\i sica, Universidad de La Laguna, C/ Astrof\'isico Francisco S\'anchez, 38205 La Laguna, Spain
             }
   \date{ }

 
  \abstract
   {The tumbling pattern of a bar is the main parameter characterising its dynamics. From numerical simulations, its evolution since bar formation is tightly linked to the dark halo in which the bar is formed through dynamical friction and angular momentum exchange. Observational measurements of the bar pattern speed with redshift can restrict models of galaxy formation and bar evolution. }
  {We aim  to determine, for the first time,  the bar pattern speed evolution with redshift based on morphological measurements.}
   {We have  selected  a  sample of  44  low inclination  ringed
galaxies from the SDSS and  COSMOS surveys covering the redshift range
0$<z<$0.8 to investigate  the evolution of the bar  pattern speed.  We
have derived  morphological ratios between the  deprojected outer ring
radius  ($R_{ring}$) and the  bar size  ($R_{bar}$). This  quantity is
related to the  parameter ${\cal R}=R_{\rm CR}/R_{bar}$ used for
classifiying  bars  in  slow  and  fast  rotators,  and  allow  us  to
investigate possible differences with redshift.}
{We obtain a similar distribution of $R$  at all redshifts.  We do not find any systematic effect that could be forcing this result.}
{The  results obtained here  are compatible with  both, the
bulk of  the bar population  ($\sim 70\%$) being fast-rotators  and no
evolution of the  pattern speed with redshift.  We  argue that if bars
are  long-lasting structures,  the results  presented here  imply that
there has not been a substantial angular momentum exchange between the
bar and  halo, as predicted by numerical  simulations. In consequence,
this  might imply  that  the discs  of  these high  surface-brightness
galaxies are maximal.}
\authorrunning{P\'erez, Aguerri \& M\'endez-Abreu}
\titlerunning{Bar pattern speed evolution over the last 7 Gyr}
 \maketitle
%

\section{Introduction}
\label{introduction}

Stellar bars  are thought to  be the main internal  mechanism driving
the dynamical and  secular evolution of disc galaxies.   They are able
to  modify the central  structure and  morphology of  galaxies, mostly
because they are responsible  for a substantial redistribution of mass
and  angular  momentum   in  the  discs  \citep[e.g.,][]{sellwood1981,sellwood1993, pfenniger1991,  athanassoula2003, debattista2006}.  In
the last  decade, numerical simulations have  addressed the importance
of  the transference  of angular  momentum between  baryonic  and dark
matter  components \citep[e.g.,][]{debattista1998,debattista2000}.   The amount
of angular momentum exchanged is related to the specific properties of the
galaxies,  such as  the  bar  mass, halo  density,  and halo  velocity
dispersion  \citep{debattista1998,athanassoula2003,sellwood2006} and  it  takes place
mainly           at           the           disc           resonances
\citep{athanassoula2003,martinezvalpuesta2006}.   Recent  works  have
shown  that  gas fraction  can  also play  an  important  role in  the
bar-halo    interplay   and   therefore    in   the    bar   evolution
\citep{bournaud2002,romanodiaz2009,villavargas2010}.   Moreover,  bars
are efficient at funneling material toward the galaxy centre and possibly
they influence the building of  the stellar bulge \citep[e.g.,][]{kormendy2004} and
the   central  black  hole   \citep[e.g.,][]{shlosman1989,  regan1999,
  corsini2003}.  Peanut/boxy bulges in galaxies are also thought to be
associated  with  bending instabilities  and  bar vertical  resonances
\citep{bureaufreeman1999, debattista2004, debattista2006,athanassoula2005,martinezvalpuesta2006, mendezabreu2008}.

As already mentioned, the bar formation  and destruction rate, as well  as the morphological
and dynamical changes suffered by  the discs during their lifetime are
strongly affected  by the  angular momentum exchange.   Therefore, the
cosmological evolution  of the  bar fraction can  also depend  on this
effect.   Observations show  that bars  in low  redshift  galaxies are
ubiquitous,  with  a  fraction  of $\sim$45\%  at  visual  wavelengths
\citep[e.g.,][]{marinova2007,   reese2007,  barazza2008,  aguerri2009,
  masters2011}   and  rising  to   $\sim$70\%  in   the  near-infrared
\citep{knapen2000,  eskridge2000,   menendezdelmestre2007}.   The  bar
fraction  depends on  morphological  type, being  lower in  lenticular
galaxies than in spirals (Marinova \& Jogee 2007; Aguerri et al. 2009; Nair \& Abraham 2010; Barway et al. 2011;
but see also Masters et al.  2011)\nocite{barway2011}.  Some recent results show that bar
fraction is  a strong function of  galaxy mass \citep{mendezabreu2010,
  nair2010} and color \citep{hoyle2011}.  In  contrast, bar fraction  is  only barely affected  by the
environment 
\citep{aguerri2009,li2009,mendezabreu2010}.  

The  evolution of the bar
fraction   with    redshift   is    still   a   matter    of   debate.
\citet{abraham1999}  found that the  fraction of  barred galaxies  at $z>0.5$
is lower than the local fraction.  However, other authors claim that this
may be the  consequence of selection effects, due  to the high angular
resolution needed to find bars (Elmegreen et al. 2004; but see van den
Bergh  2002)\nocite{elmegreen2004,vanderbergh2002}.  To  deal with  the angular  resolution  problem, several
studies have carried  out this analysis using the  Advanced Camera for
Surveys (ACS) on the Hubble  Space Telescope (HST). Thus, Elmegreen et
al.   (2004) and  Jogee et  al.  (2004)  found the  same  bar fraction
($\sim$40\%)  at   redshift  $z=1.1$  as  in   the  local  Universe,
suggesting that the  bar dissolution cannot be common  during a Hubble
time  unless  the  bar  formation   rate  is comparable  to  the  bar
destruction rate.  On the contrary,  Sheth et al.  (2008)\nocite{sheth2008}, in a recent
study  using   images  from   the   Cosmological  Evolution   Survey
\citep[COSMOS;][]{scoville2007}  and   using  a  larger   sample  than
previous studies, found that the bar fraction at $z=0.84$ is one-third
of the present-day  value.  They also found a  much stronger evolution
for low mass galaxies and  late-type morphological types.  Part of the
differences may  be due  to  the selection  effects and  other
systematic effects that still need to be investigated further.

In any case, these results show that bars have been common structural components of the discs of galaxies during the last 8 Gyrs. The study of their origin and evolution could be crucial for understanding the galaxy evolution since $z=1$. 
This study can be done by analysing three parameters that characterise the bars: length,  strength and
pattern  speed.    Several  methods and techniques  have been proposed in  order to
measure  these bar  parameters.   The bar  length  has been  obtained
directly  by visual inspection  on galaxy  images \citep{kormendy1979,
  martin1995, mendezabreu2010, nair2010,  masters2011},  searching for
the  maximum ellipticity of  the galaxy  isophotes \citep{wozniak1995,
  laine2002,  marinova2007, aguerri2009},  locating variations  of the
isophotal  position   angle  \citep{sheth2003,  erwin2005},  analyzing
Fourier  moments \citep{quillen1994,  aguerri2000,  aguerri2003}, or  by
photometric  decomposition  of   the  surface-brightness  profiles  of
galaxies \citep{prieto1997,  prieto2001, aguerri2005, laurikainen2005,
  gadotti2008, weinzirl2009, gadotti2011}. The  resulting studies  reported that the  typical bar
length is  about 3-4 kpc, and  strongly correlates  with the disc
scale-length  \citep{aguerri2005,perez2005,erwin2005,marinova2007,laurikainen2007}. Bar
length  is  also a  function  of  galaxy  size, morphology  and  color
\citep{aguerri2009, hoyle2011}.  

The  bar strength has been determined
by  measuring  bar  torques  \citep{buta2001},  isophotal  ellipticity
\citep{martinet1997, aguerri1999, whyte2002, marinova2007}, or Fourier
modes       \citep{otha1990,       aguerri2000,       laurikainen2005,
  athanassoula2002b}.   This parameter  depends on  galaxy morphology.
Bars  in lenticular  galaxies  are generally  weaker  than in  spirals
(Das et al. 2003; Laurikainen et  al. 2007; Barazza et al. 2008; Aguerri et  a. 2009; Buta et  al. 2010)\nocite{das2003,buta2010}.

The  bar  pattern  speed,  $\Omega_{\rm  b}$,  is  the  main  kinematic
observable  and describes  the  dynamics of  the  bar.  This  tumbling
pattern determines the  position of the resonances in  the disc and it
is  most usefully  parametrised  by a  distance independent  parameter
${\cal   R}=R_{\rm  CR}/R_{\rm   bar}$,  where   $R_{\rm  CR}$   is  the
Lagrangian/corotation radius, where  the gravitational and centrifugal
forces cancel  out in the rest frame  of the bar, and  $R_{\rm bar}$ is the
bar  semi-major  axis.   Therefore,  bars  that  end  near  corotation
(1$<{\cal  R}<$1.4) are  considered fast,  while shorter  bars (${\cal
  R}>$1.4) are commonly called slow. If ${\cal R}<1.0$ then orbits are
elongated perpendicular  to the bar,  and self consistent  bars cannot
exist  in  this  regime  \citep{contopoulos1980}.  The  most  reliable
method for obtaining  the location of corotation was  that proposed by
Tremaine \& Weinberg  (1984, hereafter TW method)\nocite{tremaine1984} which  uses a set of
simple kinematic measurements to derive the bar pattern speed assuming
that the tracer obeys the continuity equation, that the discs are flat
and  that there  is one  well  defined pattern  speed. However,  large
integration times are required  in medium-size telescopes to reach the
high signal-to-noise required to apply  the TW method. This limits its
application to a small number of candidates.  Despite the difficulties
in obtaining bar pattern speeds, a reasonable number of nearby galaxies
have   been    investigated   \citep{merrifield1995,   debattista2002,
  aguerri2003,   corsini2007}   finding  that   all   bars  end   near
corotation. Some  of these assumptions are not applicable for galaxies with nested bars,
and  there is  now a  simple extension  of the  TW method  to multiple
pattern speeds \citep{maciejwski2006,  corsini2003, meidt2009} and the
fact that some authors have shown that the TW method can be applied to
CO  \citep{rand2004,  zimmer2004}  and  H$_{\alpha}$  velocity  fields
\citep{hernandez2005,     emsellem2006,     fathi2007,     chemin2009,
  gabbasov2009, fathi2009} opens a new window to these studies.

Some indirect  ways to  derive the bar  pattern speed  include methods
based on numerical modelling: generating either self-consistent models
or  models  using  potentials  derived from  the  light  distributions
\citep{duval1983,  lindblad1996,   laine1998,  weiner2001,  perez2004,
  zanmar2008} and  then matching  numerical experiments  with the
observed velocity fields; or  by matching numerical simulations to the
galaxy    morphology   \citep{hunter1988,    england1989,   laine1998,
  aguerri2001, rautiainen2005}.  Other  indirect methods to derive the
bar  pattern  speed  include  identifying morphological  or  kinematic
features    with   resonances:   using    a   variety    of   features
\citep{elmegreen1990};      the      shape      of     dust      lanes
\citep{athanassoula1992}; the  sign inversion of  the radial streaming
motion  across  corotation  \citep{canzian1993};  rings  as  resonance
indicators   \citep{buta1986,   buta1995};   phase-shift  between   the
potential  and density  wave patterns  \citep{zhang2007};  location of
minimum of star  formation \citep{cepa1990,aguerri2000}; or comparison
of the behaviour of the  phase Fourier angle in blue and near-infrared
images  \citep{puerari1997,aguerri1998}.  Although  possibly  the most
accurate  indirect  method to  calculate  pattern  speeds  is the  the
comparison  of   gas  velocities   to  those  obtained   in  numerical
simulations   that  use   a   potential  obtained   from  optical   or
near-infrared light,  it is also very  time consuming and  can only be
applied to a relatively small number of objects.

The technique to  determine the bar pattern speed  based on connecting
the location  of rings  to orbital resonances  was introduced  by Buta
(1986). It is based on the  theoretical work presented by Schwarz in a
series   of  papers  \citep{schwarz1981, schwarz1984a, schwarz1984b}
showing how  these ring structures appear near  the dynamical Lindblad
resonances due  to a bar--like  perturbation.  To directly  apply this
method to  find the specific value  of the pattern speed  not only the
location of the  ring and the association to  a resonance is required,
but some  kinematic information is  also needed.  However, we  can use
the  ${\cal R}$ parametrisation  of the  bar introduced  previously, and
determine the ratio between the outer ring radius (linked to the outer
Lindblad  resonance, OLR)  and the  bar length.  In this  way,  we can
indirectly  determine, not  the pattern  speed, but  whether  the bars
measured are  in the slow or fast  regime.  

The bar parameters discussed above have been analysed in local galaxy samples. There are no previous studies in the literature about the evolution of the length, strength and pattern speed of bars. In this article, we study  for the
first time, the dynamical evolution  of bars with redshift, going from
the local Universe to $z\sim0.8$. We use a well selected sample of barred galaxies with
outer rings  to exploit the  power of this  method. The study of the dynamical evolution of bars is
critical to  constrain the angular momentum exchange between the disc and the halo and their subsequent evolution.
Weinberg (1985)\nocite{weinberg1985} predicted that  a bar would
lose  angular momentum  due  to  a massive  dark  matter halo  through
dynamical friction, slowing down  in the process.  This prediction was
further  confirmed  in  numerical  simulations  \citep{debattista1998,
  debattista2000,  athanassoula2003,  sellwood2006}  where they  found
that  bars are  slowed efficiently  if a  substantial density  of dark
matter is present in the region of the bar.  On the other hand, if the
mass  distribution is  dominated by  the  stellar disc,  then the  bar
remains rapidly  rotating for a long  time. We show in  this work that
bars do not  show a systematic change in their  dynamical state in the
last $\sim7$ Gyrs.
 
The article is  organised as follows: we present  the sample selection
and  morphology  discussion  in  Sect.~\ref{sample}. We  describe  the
method followed to measure the  ring and bar radius in Sect.~\ref{olr}.   The results are  presented in  Sect.~\ref{results} and  we discuss
their implications in Sect.~\ref{discussion}. Conclusions are provided
in Sect.~\ref{conclusions}. Throughout the paper the cosmological parameters used are: $H_{0}=70$ km s$^{-1}$ Mpc$^{-1}$, $\Omega_{\Lambda}=0.7$, and $\Omega_{m}=0.3$.
 

\section{Sample selection}
\label{sample}
 
The galaxy  samples studied  in this article  were extracted  from two
different  surveys: low redshift  galaxies were  taken from  the Sloan
Digital  Sky Survey  (SDSS;  $0.01 <  z  < 0.04$),  and high  redshift
galaxies were selected  from COSMOS ($0.125 < z  < 0.75$). 

Two caveats must be discussed  before the samples are described in
detail:  first, it is  worth noticing  that the  galaxy samples  are not
meant to be complete in  any sense, however, the selection criteria make the two samples fully comparable. Second, in Aguerri et
al. (2009) we studied how the resolution of the SDSS images can affect
our detection of bars.  We worked out, using artificial galaxies, that
the shortest bars that we are able to resolve have a length of $\sim9$
pixels.  Considering  a mean  PSF in  our SDSS images  with a  FWHM of
1\farcs09 (2.77 pixel),  we conclude that we resolve  bars larger than
$\sim3\times$  FWHM, or  equivalently, $\sim0.5$  kpc at  $z=0.01$ and
$\sim2$ kpc at $z=0.04$.  The COSMOS sample was selected using the ACS
data in the  F814W filter.  The images were  processed to a resolution
of  0\farcs05 pixel$^{-1}$  with an  averaged PSF  FWHM  of 0\farcs097
\citep{scoville2007,    koekemoer2007}.    Based   on    the   previous
considerations,  we will  resolve bars  larger  than $\sim3\times$FWHM
which corresponds to  $\sim0.6$ kpc at $z=0.125$ and  $\sim2.2$ kpc at
$z=0.75$, matching  perfectly the SDSS  spatial resolution in  the low
redshift range.
 
\subsection{Outer      Ring     Morphological Classification}
\label{OLRmorph}

\begin{figure*}
\begin{center}
\resizebox{1.\textwidth}{!}{\includegraphics[angle=0]{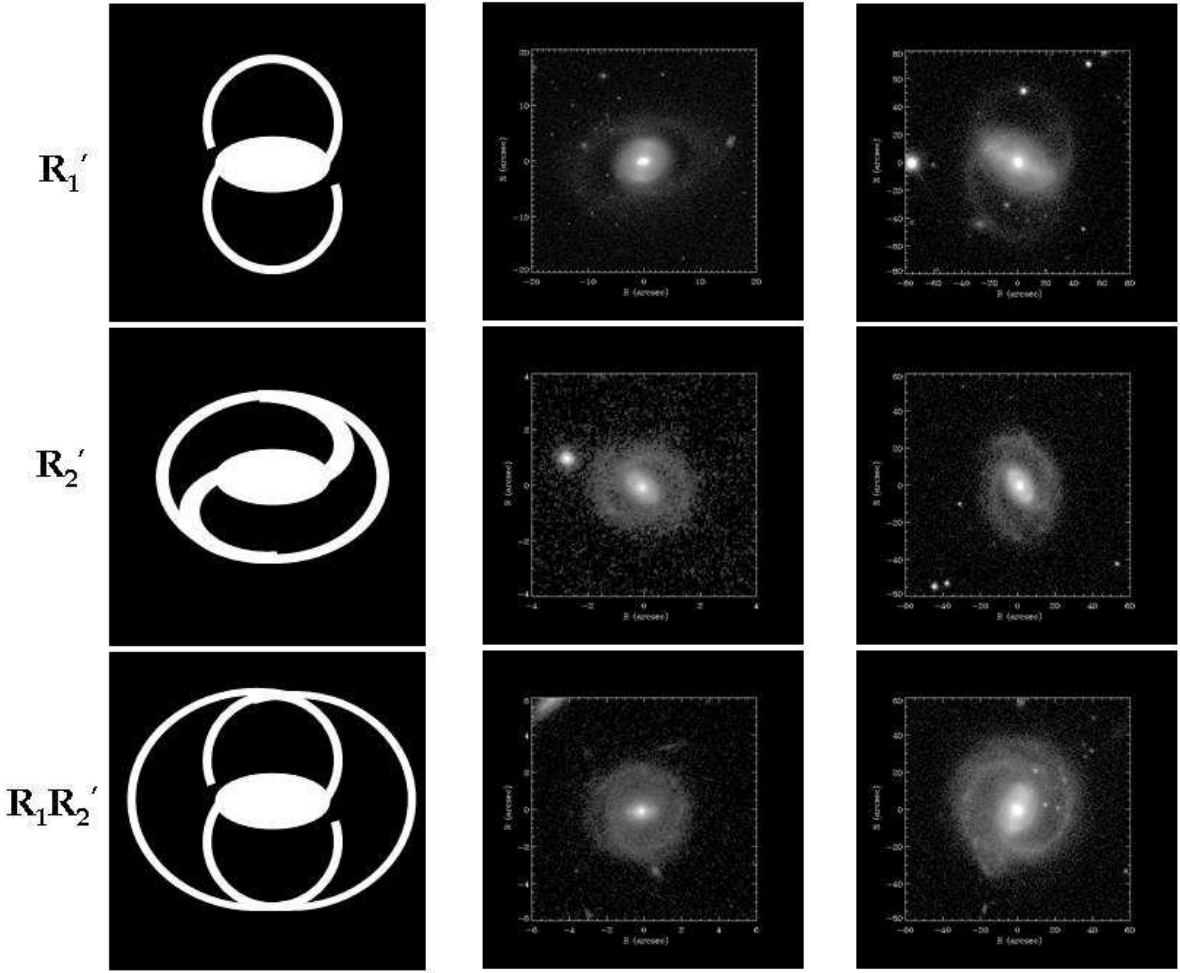}}
\caption{Left column: schematic view of the different ring types in barred galaxies (see Sect. 2.1). Ring type examples of barred galaxies at high (middle column), and low (right column) redshifts.  \label{fig:ringtype}}
\end{center}
\end{figure*}

The  ring morphological classification used in this study is  based on the
work of \cite{butacrocker1991}.  They divide the outer rings in three
main morphological classes resembling the rings developed in
numerical  simulations  near the  OLR  \citep{schwarz1981}.  The  first
class, called  R$_{1}'$,  is  characterised  by a  180$^{\circ}$~winding of  the spiral arms  with respect to  the ends of a  bar.  The
second  type  is known  as  an  R$_{2}'$ ring.   It  is  defined by  a
270$^{\circ}$~winding of the outer  arms with respect to the bar ends,
so that  in two  opposing quadrants the  arm pattern is  doubled.  The
R$_{1}'$ and R$_{2}'$ morphologies were predicted by Schwarz (1981) as
the kind of patterns that would  be expected near the OLR in a barred
galaxy.  The third  class is referred to in Buta  \& Crocker (1991) as
the R$_{1}$R$_{2}'$  morphology, where the  outer arms break  not from
the ends of the bar, but  from an R$_{1}'$-type ring.  The existence of
this combined type, which may be linked to the population of both main
families of  OLR periodic orbits \citep{schwarz1981},  provides some of
the clearest  evidence of  the OLR in  barred galaxy  morphology. Some
examples of this classification, taken from our sample of low and high
redshift galaxies,  are shown  in Fig.~\ref{fig:ringtype}.  Buta et
al. (1995) derived the  distribution of intrinsic axis  ratios for
the outer  rings using  the Catalog of  Southern Ringed  Galaxies.  They
found that  outer rings present  in barred galaxies  are intrinsically
elliptical  with  an  axis   ratio  $\sim0.82\pm0.07$,  and  that  the
intrinsic ellipticity  varies from the  R$_{1}'$ ($\sim0.74\pm0.08$) to
the  R$_{2}'$ ($\sim0.87\pm0.08$).   The intrinsic  shape of  the rings
plays an  important role when  deprojecting distances such as  the bar
length and the ring  radius, thus, more intrinsically elliptical rings
will increase  the uncertainties in  the measurements.  We  decided to
remove from  our samples  the R$_{1}'$ type  of rings, and  keep only the R$_{2}'$ types since  they are intrinsically rounder. In fact,
their  intrinsic  shape is  very  similar  to  that of  typical  discs
\citep{fasano1993,ryden2004}.

\subsection{Low redshift}
\label{lowz}

The  barred ringed  galaxies at  low redshift  were obtained  from the
galaxy sample  analysed in  Aguerri et al.   (2009).  They  selected a
volume limited sample of  galaxies from the spectroscopic catalogue of
the SDSS Data Release 5 (SDSS-DR5, Adelman-McCarthy et al. 2007)\nocite{adelman2007}. This
sample  covers the redshift  range $0.01<z<0.04$,  down to  an absolute
magnitude  of $M_{r}<-20$,  and with  low inclination  $i<60^{\circ}$. The
full sample consist  of 3060 galaxies with a  morphological mix of 26\%
ellipticals, 29\% lenticulars, 20\%  early type spirals, and 25\% late
type spirals. Galaxies were  classified in barred and unbarred systems
by searching for  absolute maxima in the ellipticity  radial profiles of
their isophotes (see Aguerri et al. 2009 for details). From the barred
sample we visually  inspected the SDSS galaxy images  in order to look
for the presence of outer rings of type R$_{2}'$.

We obtained  a total of 18  barred galaxies with  suitable outer rings
features. Table \ref{samplesdss.tab} shows  the main properties of the
bars and rings features measured in these galaxies.

\subsection{High redshift}

As for the low  redshift sample, we have chosen a
number  of low  inclination galaxies  from  the third  release of  the
COSMOS HST survey \citep{scoville2007}.  We first downloaded all the 81
image  tiles from  the  COSMOS/ACS fields,  observed  using the  F814W
($I$-band)   filter,   from   the   Multimission  Archive   at   STScI
(MAST\footnote{Based  on observations  made with  the  NASA/ESA Hubble
  Space  Telescope,  obtained  from  the  data archive  at  the  Space
  Telescope Science Institute. STScI is operated by the Association of
  Universities for Research in Astronomy, Inc. under NASA contract NAS
  5-26555.})  for visual inspection.  These observations cover $\sim$ 2
deg$^{2}$  with a  pixel scale  (for the  drizzled data)  of 0\farcs05
pixel$^{-1}$.

We  visually scanned  the COSMOS  fields  to look  for clearly  ringed
barred galaxies.  After a  preliminary list was created, we correlated
the  positions with the  spectroscopic redshifts  from the  Very Large
Telescope  (VLT) and  Magellan COSMOS  spectroscopic  surveys (zCOSMOS
Survey;  Lilly et  al.   2007) to  search  for candidates  with
reliable redshifts.   We searched  also for photometric  redshifts for
the remaining candidates.  We used photometric redshifts determined by
Faure  et  al.   (2008)   using  the  Le  Phare  photometric  redshift
estimation  code (Ilbert  et al.   2006), details  concerning the
multi-wavelength   photometry   can    be   found   in   Mobasher   et
al. (2007). Faure et al. (2008)\nocite{faure2008,mobasher2007,ilbert2006,lilly2007} used 1095 spectroscopic redshifts from
the zCOSMOS Survey  (Lilly et al. 2006) to  calibrate the ground-based
photometric  zero points.  Using eight  bands, this
method  achieves a  photometric redshift  accuracy  of $\sigma_{\Delta
  z}$/(1+$z_{s}$)=0.031.

\begin{figure}
\begin{center}
\resizebox{0.5\textwidth}{!}{\includegraphics[angle=0]{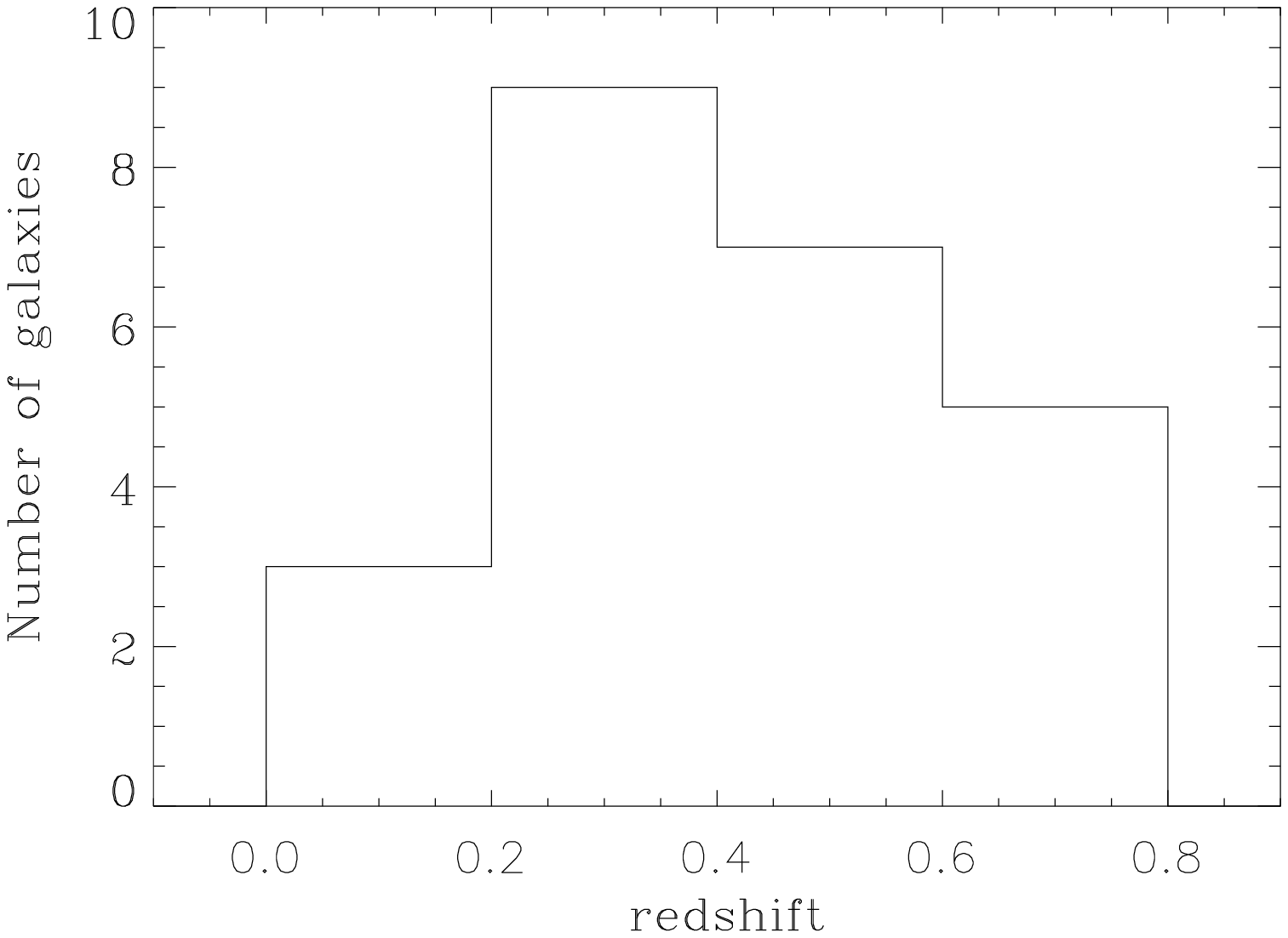}}
\caption{Redshift  distribution for the  high redshift  sample. \label{fig:histo}}
\end{center}
\end{figure}

From  this  sample,  only  galaxies  showing type  R$_{2}'$  ring  were
included in the  sample. Finally the high redshift  sample consists of
26 galaxies.  The  main properties of the bars  and rings features are
shown in Table~\ref{samplez.tab}.  The sample covers the redshift range
$0.125<z<0.75$.     Figure   \ref{fig:histo}   shows    the   redshift
distribution.

\begin{table*}[!ht]
\begin{center}
\caption{General properties of the SDSS sample.\label{samplesdss.tab}}
\begin{tabular}{c c c c c c c c}     
\hline\hline
Name  & R$_{\rm bar-min}$ &  R$_{\rm bar-max}$ &  PA$_{\rm bar}$ &$\epsilon_{\rm ring}$  & PA$_{\rm ring} $  &  R$_{\rm ring}$  &  $z$\\
      &    (kpc)     &    (kpc)     &    (degrees)&            & (degrees)    & (kpc)       &\\
  (1) &      (2)     &     (3)       &     (4)    &    (5)     &     (6)      &    (7)      & (8) \\
\hline
SDSSJ104924.86-002547.5  &     6.6  &    5.1    &   121.6$\pm$0.9  &    0.402$\pm$0.004  &   142.2$\pm$0.8  &    13.0  &       0.039 \\
SDSSJ102543.29+393846.9  &     7.3  &    5.1    &   153.4$\pm$0.2  &    0.075$\pm$0.005  &    74.8$\pm$2.3  &    13.0  &       0.023 \\
SDSSJ122529.23+471623.4  &     5.7  &    4.7    &    11.6$\pm$0.1  &    0.174$\pm$0.005  &    30.1$\pm$14.3 &    10.8  &       0.025 \\
SDSSJ130235.73+411924.1  &     4.8  &    3.7    &   165.4$\pm$0.7  &    0.186$\pm$0.005  &    45.4$\pm$1.5  &     9.9  &       0.028 \\
SDSSJ120732.62+324846.7  &     7.1  &    4.7    &    54.4$\pm$0.5  &    0.084$\pm$0.007  &    74.1$\pm$20.0 &    11.6  &       0.026 \\
SDSSJ133259.13+321913.6  &     5.7  &    3.8    &   177.3$\pm$0.3  &    0.044$\pm$0.007  &   106.7$\pm$8.6  &     8.7  &       0.035 \\
SDSSJ012858.63-005656.3  &     8.0  &    6.3    &     6.2$\pm$0.3  &    0.341$\pm$0.002  &    83.6$\pm$0.4  &    16.4  &       0.018 \\
SDSSJ083220.43+412132.0  &     3.5  &    2.4    &    52.8$\pm$0.5  &    0.048$\pm$0.005  &   111.3$\pm$10.9 &     7.9  &       0.025 \\
SDSSJ083630.84+040215.6  &     6.6  &    4.5    &    94.2$\pm$0.2  &    0.164$\pm$0.009  &    37.0$\pm$3.8  &    12.7  &       0.029 \\
SDSSJ091426.23+360644.1  &     6.3  &    5.0    &   164.4$\pm$0.0  &    0.163$\pm$0.031  &   156.1$\pm$1.0  &    10.1  &       0.022 \\
SDSSJ123234.57+492312.2  &     5.3  &    3.4    &   176.1$\pm$0.5  &    0.074$\pm$0.012  &   114.1$\pm$10.7 &     8.7  &       0.040 \\
SDSSJ142412.12+350846.0  &     3.6  &    2.7    &    45.3$\pm$1.0  &    0.238$\pm$0.006  &    52.9$\pm$26.7 &     9.3  &       0.029 \\
SDSSJ153619.30+493428.3  &     5.3  &    2.9    &    17.8$\pm$2.1  &    0.211$\pm$0.006  &    73.1$\pm$0.7  &     9.7  &       0.038 \\
SDSSJ160331.62+492017.3  &     8.7  &    5.2    &    70.9$\pm$0.4  &    0.255$\pm$0.005  &    32.3$\pm$7.2  &    16.9  &       0.020 \\
SDSSJ172721.89+593837.6  &     5.9  &    4.0    &   171.3$\pm$0.1  &    0.202$\pm$0.004  &   148.8$\pm$1.1  &    11.4  &       0.028 \\
SDSSJ123313.69+121449.2  &     3.9  &    3.1    &   118.5$\pm$0.9  &    0.023$\pm$0.004  &    65.3$\pm$11.5 &     6.9  &       0.026 \\
SDSSJ120609.11-025653.2  &     5.3  &    3.9    &    61.5$\pm$0.7  &    0.149$\pm$0.009  &    17.3$\pm$4.6  &    11.2  &       0.026 \\
SDSSJ111044.88+043039.0  &     6.8  &    5.0    &    97.6$\pm$0.2  &    0.062$\pm$0.007  &    64.6$\pm$9.2  &    12.5  &       0.029 \\
\hline
\end{tabular}
\smallskip
\begin{minipage}{140mm} 
NOTE. Col. (1): Galaxy name from SDSS; Col. (2): bar radius calculated
using the  position of the  minimum ellipticity; Col. (3):  bar radius
calculated using  the position of  the maximum ellipticity;  Col. (4):
position  angle of  the bar;  Col.  (5): ring  ellipticity; Col.  (6):
position  angle  of  the  ring;  Col.  (7):  ring  radius;  Col.  (8):
spectroscopic redshift from SDSS
\end{minipage} 
\end{center}
\end{table*}

\begin{table*}[!ht]
\begin{center}
\caption{General properties of the COSMOS sample\label{samplez.tab}}
\begin{tabular}{c c c c c c c c}     
\hline\hline
Name & R$_{\rm bar-min}$ &  R$_{\rm bar-max}$ &  PA$_{\rm bar}$ & $\epsilon_{\rm ring}$ & PA$_{\rm ring}$ &  R$_{\rm ring}$  &  $z$\\
      &    (kpc)     &    (kpc)     &    (degrees)&            & (degrees)    & (kpc)       &\\
  (1) &      (2)     &     (3)       &     (4)    &    (5)     &     (6)      &    (7)      & (8) \\
\hline
812947     &     	5.4  &  4.1 &  76.3$\pm$0.1  & 0.068$\pm$0.010  & 120.5$\pm$14.8 &  9.8   &  0.125~($s$)\\
816960     & 		5.5  &  3.4 &  174.8$\pm$1.2 & 0.059$\pm$0.035  &  85.2$\pm$22.6 &  7.5   &  0.311~($s$)\\
J095928.30+020109.0  &  5.3  &  3.8 &  63.2$\pm$0.6  & 0.088$\pm$0.045  & 141.3$\pm$41.8 &  9.1   &  0.530~($p$)\\
817887      & 		5.6  &  4.6 &  116.2$\pm$2.8 & 0.056$\pm$0.028  &  88.4$\pm$17.5 & 15.1   &  0.672~($s$)\\
823705      &    	3.6  &  2.7 &   46.4$\pm$0.7 & 0.165$\pm$0.028  &  74.4$\pm$17.2 &  6.3   &  0.491~($s$)\\
824759      &    	3.7  &  2.9 &  136.1$\pm$1.8 & 0.161$\pm$0.023  & 158.1$\pm$ 7.0 &  7.0   &  0.751~($s$)\\
825492      &   	5.1  &  3.6 &  108.4$\pm$2.4 & 0.152$\pm$0.057  & 108.1$\pm$22.2 &  8.7   &  0.736~($s$)\\
833039      &  		4.3  &  3.3 &  100.6$\pm$1.1 & 0.203$\pm$0.056  &  32.8$\pm$ 4.8 &  8.8   &  0.360~($s$)\\
J100233.98+022524.3  &  4.7  &  2.2 &  172.7$\pm$0.9 & 0.167$\pm$0.015  & 63.1$\pm$ 4.4  &  9.4   &  0.720~($p$)\\
J095938.81+020658.7  &  10.3 &  7.6 &  111.7$\pm$0.4 & 0.071$\pm$0.035  & 149.6$\pm$75.9 & 16.6   &  0.409~($p$)\\
J095935.08+020127.2  &  4.3  &  3.0 &  176.8$\pm$79.2& 0.158$\pm$0.044  &  26.3$\pm$36.9 & 10.3   &  0.357~($p$)\\
J100204.95+022739.7  &  5.8  &  3.7 &  124.4$\pm$1.0 & 0.116$\pm$0.019  & 122.7$\pm$ 6.6 & 10.5   &  0.507~($p$)\\
841055      &		5.7  &  3.4 &  134.9$\pm$0.6 & 0.188$\pm$0.020  & 178.0$\pm$56.2 & 13.7   &  0.376~($s$)\\
J095759.45+022810.5  &  4.4  &  2.6 &  20.6$\pm$0.9  & 0.042$\pm$0.013  & 131.0$\pm$23.0 &  9.2   &  0.119~($s$)\\
851598      &     	6.4  &  4.4 &  9.6$\pm$0.5   & 0.080$\pm$0.047  & 102.2$\pm$30.7 & 10.0   &  0.346~($s$)\\
852495      & 		6.8  &  5.7 &  96.7 $\pm$3.1 & 0.265$\pm$0.015  &  21.9$\pm$ 1.5 & 12.2   &  0.705~($s$)\\
852636      &   	7.1  &  4.9 &  0.7$\pm$80.0  & 0.040$\pm$0.026  & 125.6$\pm$50.7 & 12.5   &  0.345~($s$)\\
852155      &	        7.7  &  5.6 &  173.0$\pm$0.3 & 0.306$\pm$0.009  & 160.4$\pm$ 0.8 & 15.3   &  0.305~($s$)\\
J100254.88+024645.8  &  4.4  &  3.2 &  18.7$\pm$1.5  & 0.065$\pm$0.036  & 128.0$\pm$45.7 &  8.5   &  0.468~($p$)\\
840577      &     	5.1  &  3.5 &  68.3$\pm$2.4  & 0.085$\pm$0.023  &  89.5$\pm$15.0 &  7.9   &  0.539~($s$)\\
838743      &  		4.4  &  2.8 &  48.7$\pm$0.5  & 0.144$\pm$0.025  &  60.0$\pm$3.8  &  8.9   &  0.126~($s$)\\
830974      & 		9.6  &  5.3 &  11.5$\pm$1.4  & 0.149$\pm$0.025  &  57.7$\pm$7.0  & 13.9   &  0.695~($s$)\\
811921      &      	4.9  &  3.1 &  25.2$\pm$1.6  & 0.068$\pm$0.027  &  89.5$\pm$30.7 &  8.2   &  0.371~($s$)\\
813153      & 		7.5  &  6.0 &  60.0$\pm$0.7  & 0.159$\pm$0.069  &  47.3$\pm$28.1 & 13.2   &  0.529~($s$)\\
831775      &		3.6  &  1.8 &  125.1$\pm$0.1 & 0.140$\pm$0.028  & 156.8$\pm$76.5 &  6.2   &  0.381~($s$)\\
J100217.12+023024.1   & 4.9  &  3.4 &  157.8$\pm$0.8 & 0.083$\pm$0.022  &  97.7$\pm$15.6 &  8.1   &  0.379~($p$)\\
\hline
\end{tabular}
\smallskip
\begin{minipage}{140mm} 
NOTE.  Col.  (1):  Galaxy  name  from COSMOS;  Col.  (2):  bar  radius
calculated using  the position of  the minimum ellipticity;  Col. (3):
bar radius  calculated using the position of  the maximum ellipticity;
Col. (4): position angle of the bar; Col.  (5): ring ellipticity; Col.
(6): position  angle of the ring;  Col.  (7): ring  radius; Col.  (8):
($s$):  spectroscopic   redshift  from  zCOSMOS;   ($p$):  photometric
redshift from Faure et al.  (2008).
\end{minipage} 
\end{center}
\end{table*}

\section{Outer ring radius and bar size definition}
\label{olr}


Our approach to quantifying the dynamical state of the bars in our sample
of ringed  galaxies is based on  the measurements of both  the bar and
ring radius. We deproject bar size in the plane of the galaxy using $i$, the galaxy  inclination, and  $\theta$, the position
angle of  the galaxy  component (bar or  ring).  We assume that  the outer ring reflects the properties of the disc,
and therefore that the ellipticity  and position angle of the ring and
disc are  the same.  In Sect.~\ref{OLRmorph} we  have discussed  that both
components are intrinsically similar  and possible differences will only
affect our  results by introducing  a large scatter.   Under this  hypothesis, the
ring radius does not need to  be deprojected, since it is measured along
the  major axis  of  the galaxy,  and  the galaxy  inclination can  be
derived simply by $i=\arccos{(1-\epsilon_{\rm ring})}$.

\subsection{Ellipse fitting } 
\label{ellipse.fitting}

The  low  redshift  sample  measurements  were derived  by  using  the
ellipticity  and position  angle  radial profiles  extracted from  the
symmetrised images.  This  approach allows us to clean  the images from
spurious  sources.   It  works  as  follows:  each  image  is  rotated
180$^{\circ}$ with respect to the galaxy centre. Then, we subtract the
rotated  frame from  the  original   one.   The  residual   image  was
sigma-clipped to identify all the pixels with a number of counts lower
than $1\sigma$, where $\sigma$ is the r.m.s. of the image background. The
value  of the deviant  pixels was  set to  zero. Finally,  the cleaned
image  was subtracted from the  original one  to  get the  symmetrised
image.  The  ellipses  were  then  fitted  to  the  isophotes  of  the
symmetrised  images of  the galaxies  using the  IRAF\footnote{IRAF is
  distributed by NOAO, which is  operated by AURA Inc., under contract
  with   the  National  Science   Foundation.}   task   {\tt  ELLIPSE}
\cite[][]{jedrzejewski1987}.
We used an iterative wrapped  procedure which runs the ellipse fitting
several times, changing the trial  values at each fit iteration, until
a good  fit at all radii  is obtained.  At each  fixed semi-major axis
length, the coordinates of the centre of the fitting ellipse were kept
fixed.  This  centre was
identified with the position of  the central intensity peak.  The trial values
for the ellipticity and position  angle were randomly chosen between 0
and  1  and between  $-90^\circ$  and  $90^\circ$, respectively.   The
fitting procedure  stopped when either convergence  was reached or
after  100  iterations.  

The  high  redshift  sample  ellipticity  and
position angle profiles were derived using the same wrapping procedure
to maximise  the goodness  of the ellipse  fitting.  However,  in this
case we preferred not to  symmetrise the images but apply a $2\times2$
pixels box  smoothing.  The image symmetrisation was  not needed since
the  sample galaxies  were not  contaminated by  other  sources within
their projected surface  and the smoothing provided better radial profiles  by improving  the signal-to-noise  ratio  of the
images.

\subsection{Bar length}
\label{bar.length}

\begin{figure*}
\begin{center}
\resizebox{1.0\textwidth}{!}{\includegraphics[angle=90]{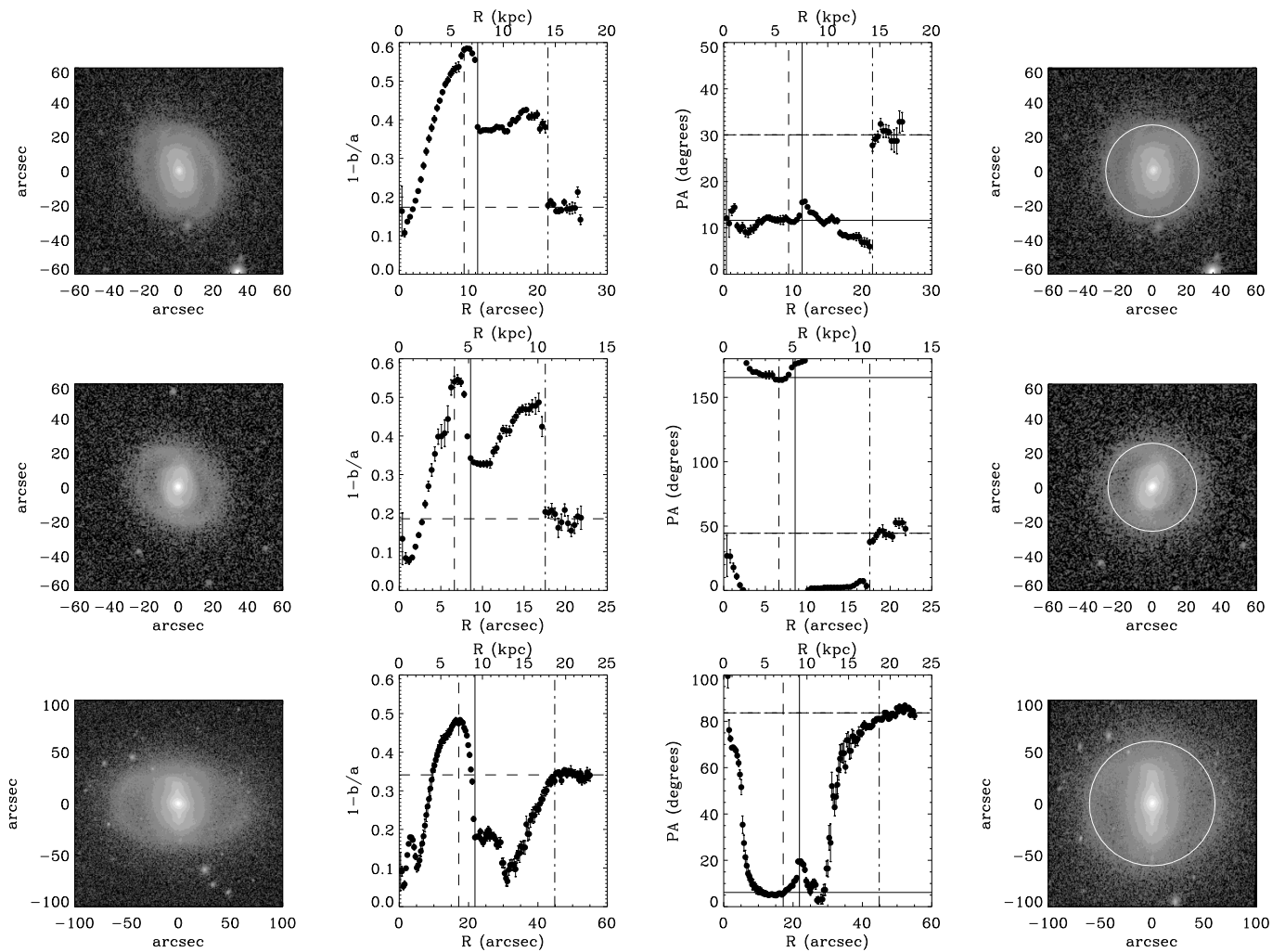}}
\caption{Three examples of the ringed galaxies selected from the SDSS data. From left to right: original $r$-band SDSS image, ellipticity isophotal radial profile, position angle radial profile of the isophotes, and deprojected image. The vertical full, dashed and dotted-dashed lines represent the R$_{\rm bar-min}$, R$_{\rm bar-max}$, and R$_{\rm ring}$, respectively. The horizontal full, dashed and dotted-dashed lines shows the PA of the bar, ellipticity of the disc and PA of the disc, respectively. The circle represented in the right-most panels has a radius equal to the measured radius of the ring.\label{fig:method1}}
\end{center}
\end{figure*}

\begin{figure*}
\begin{center}
\resizebox{1.0\textwidth}{!}{\includegraphics[angle=90]{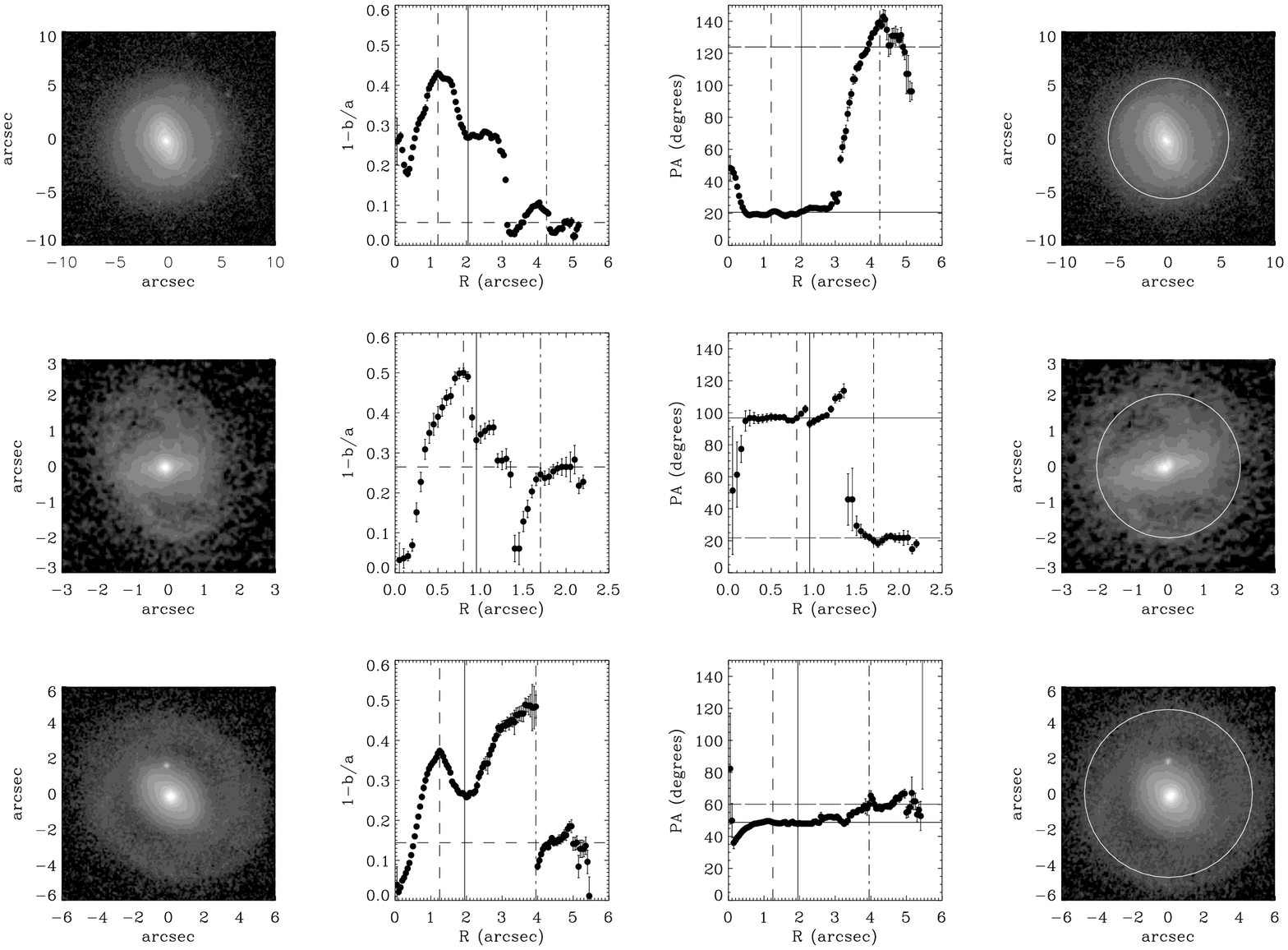}}
\caption{Three examples of the ringed galaxies selected from the COSMOS data. From left to right: original r-band COSMOS image, ellipticity isophotal radial profile, position angle radial profile of the isophotes, and deprojected image. The vertical full, dashed and dotted-dashed lines represent the R$_{\rm bar-min}$, R$_{\rm bar-max}$, and R$_{\rm ring}$, respectively. The horizontal full, dashed and dotted-dashed lines shows the PA of the bar, ellipticity of the disc and PA of the disc, respectively. The circle represented in the right-most panels has a radius equal to the measured radius of the ring.\label{fig:method2}}
\end{center}
\end{figure*}

The ellipticity radial  profile of a typical barred  galaxy reaches a minimum
at the centre, because of  either seeing effects or a spherical bulge.
Then, it usually shows a global increase to a local maximum, and
then  suddenly decreases towards  a minimum at  the location
where the  isophotes become  axisymmetric in the  face-on case.
The position angle profile  is also characteristic in barred galaxies,
being constant  in the bar region  and then changing to  fit the outer
disc orientation \citep[e.g.,][]{wozniak1995, aguerri2000b}.   These characteristic profiles are  produced by the
shape   and   orientation  of   the   stellar   orbits   of  the   bar
\cite[see][]{contopoulos1989,  athanassoula1992}.   Different  methods
have been used to measure the  bar length based on the ellipticity and
position   angle    radial   profiles   \cite[see][]{athanassoula2002,micheldansac2006}.  However, the solution is always ambiguous and it can lead to misleading results.  In order to remove these
uncertainties we  decided to  measure the bar  length as  the midpoint
between the radius of the  maximum and minimum ellipticity.  These two
different measurements  of the bar length represent  the extreme cases
\citep{micheldansac2006} and therefore they represent an upper limit of
our errors in the bar radius measurements.  We preferred this
solution not to bias our  conclusions.  The position angle of the bar,
which  is needed  to deproject  the bar  length, was  measured  at the
position of  the maximum ellipticity,  so we avoid problems  related to
position angle  variation in the  bar-disc region. An example  of this
method  applied  to three  of  our low and high redshift  galaxies is shown  in
Figs.~\ref{fig:method1}, \ref{fig:method2}, respectively.

\subsection{Ring radius}

As for the bar  component, the ring radius was derived based
on the  ellipticity and  position angle radial  profiles. In  the ring
region,  we expect  that  the ellipticity  and  position angle  radial
profiles  will  remain constant  due  to  the  stellar  orbits  in  the
ring. Therefore, we identify the  region of the profile where the ring
is present and  we measured the ring radius as  the position where the
ellipticity and position angle become constants.  The ring ellipticity
and position  angle also needed  for deprojecting, were derived  as a
mean of  these constant  values.  The error in the ring radius has been calculated by comparing the estimated ring radius with the radius at which the ellipticity varies more than 3 times the standard deviation of the disc ellipticity. Figures~\ref{appendix1} and~\ref{appendix2} from the Appendix show all the galaxies with the ring radius overplotted.
\section{Results}
\label{results}

Table~\ref{samplesdss.tab} and Table~\ref{samplez.tab} shows the obtained parameters for the ring radius, ellipticity, position angle and the bar semi-major axis, as derived in Sect.~\ref{ellipse.fitting}.  Most galaxy inclinations lie below $i<40^{\circ}$.  The bar size range, using the  maximum ellipticity, covers from 2.5 to 6.3 kpc. Most of the bars in the local Universe (about 70$\%$, see Aguerri et al. 2009) are within this bar size range. Similar values of the bar size range are found for our high redshift galaxies. The mean bar radius of our low and high redshift galaxies are 4.5$\pm$1.04 and 3.5$\pm$1.33 kpc, respectively. This means that within the errors both galaxy samples have similar bars according to their lengths and similar to local samples of barred galaxies (see Aguerri et al. 2009). The average bar size, using the minimum ellipticity, for our low and high redshift galaxies are also similar: 5.9$\pm$1.43 and 5.3$\pm$1.70 kpc, respectively. Thus, both samples of galaxies show similar bar sizes independent of the method used for determining the bar length.  It has been argued \citep{micheldansac2006} that the sizes calculated using the minimum of the ellipticity correlate well with the position of corotation, giving a more physically significant size than measurements obtained with the maximum of the ellipticity, which clearly underestimates the true bar size. To avoid problems related to the bar size calculation, as explained in Sect.~\ref{bar.length}, we have opted for using the mid-point  and to take into account the values of R$_{\rm bar}$ using both methods to obtain the errors. 

We have determined the strength of the bars for the low and high redshift galaxies by using the maximum ellipticity of the bar (see Aguerri et al. 2009). Both samples cover the same range of bar strengths. Thus, the mean values of the bar strength of our low and high redshift samples are: 0.20$\pm$0.07, and 0.17$\pm$0.05. These values are similar to the mean strength of bars in the local Universe (0.20$\pm$0.07; see Aguerri et al. 2009). We can conclude that according to the size and strength of the bars, our low and high redshift galaxy samples have similar bars as those found in a complete local  sample of barred galaxies (see Aguerri et al. 2009).

To determine whether our galaxies are in the {\it fast} or {\it slow}  range (see Sect.\ref{introduction}) we define the ratio ${\cal   R}_{\rm ring}$= R$_{\rm ring}$/R$_{\rm bar}$, where  R$_{\rm ring}$ is the ring radius and R$_{bar}$, is the bar semi-major axis, as characterised in Sect.~\ref{olr}. Because we cover this ratio for galaxies with redshifts between $0 < z < 0.8$, we can study possible changes of this ratio with redshift. Figure~\ref{fig:rbar.rolr} shows the distribution of  ring radii (R$_{\rm ring}$) vs. the
 bar semi-major axis  (R$_{\rm bar}$) for the whole sample. We consider a fast bar those bars for which the R$_{\rm CR}$/R$_{\rm bar}$ ratio lies between 1.0 and 1.4. This ratio has been plotted in Fig.~\ref{fig:rbar.rolr} for both values and is calculated using linear resonance theory and a flat rotation curve (Athanassoula et al. 1982), in this case the position of the OLR (i.e., the ring radius) and the CR are related in the following way:
 
\begin{equation}
(\frac{R_{\rm ring}}{R_{\rm CR}})^\delta= 1+(1-\frac{1}{2}\delta)^{1/2}
\label{eq:depro}
\end{equation}
 where $\delta$ lies between 0.7 and 1.0 for early type discs, see Athanassoula et al. (1982)\nocite{athanassoula1982}. We take $\delta$=1.0 in Fig.~\ref{fig:rbar.rolr} for simplicity but this choice does not alter the results. 
 It is clear from Fig.~\ref{fig:rbar.rolr} that all the galaxies, independent of their redshift bin, fall into the fast-bar category. 
 
We have investigated the influence of the inclination of the galaxies on this result. Thus, the average values for ${\cal   R}_{\rm ring}$ at different inclinations are the following: i) for the low redshift sample and  b/a $>$ 0.9, the average ${\cal   R}_{\rm ring}$ = 0.51$\pm$ 0.06, for b/a $<$ 0.9, the average ${\cal   R}_{\rm ring}$ = 0.50 $\pm$ 0.08 ii) for the high-redshift sample; for b/a $>$ 0.9, the average  ${\cal   R}_{\rm ring}$ =  0.55 $\pm$ 0.09 and for b/a $<$ 0.9, the average  ${\cal   R}_{\rm ring}$ = 0.52 $\pm$ 0.12. All the values, independently of redshift and inclination are comparable; and therefore, we do not see changes of this ratio with redshift.

\begin{figure*}
\begin{center}
\resizebox{0.7\textwidth}{!}{\includegraphics[angle=0]{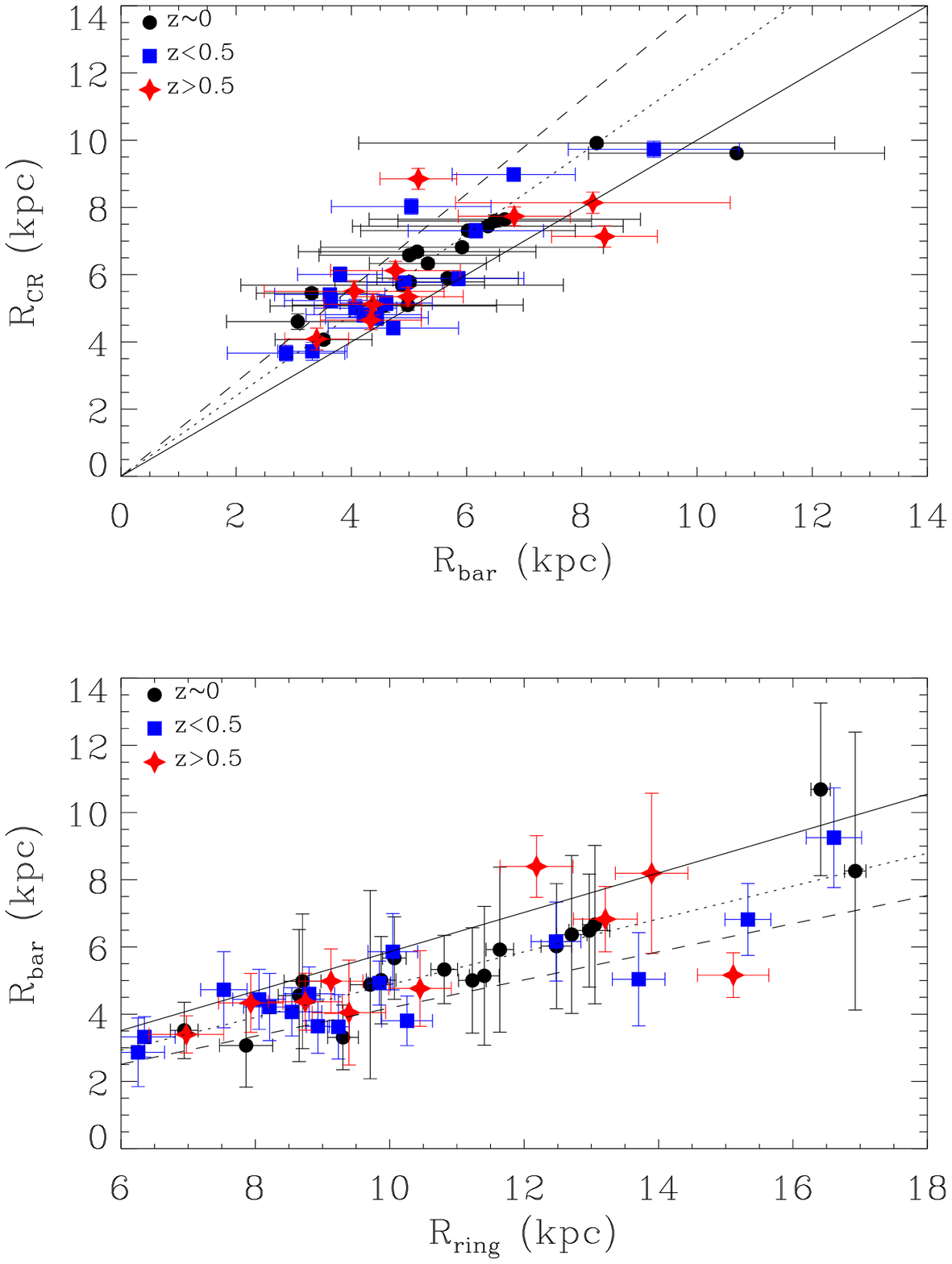}}
\caption{Top panel: bar semi-major  axis vs. corotation radius for both high and low redshift sample. The sample has been divided in three redshift bins; z$\approx$0 (solid black circles), z~$<$ 0.5 (solid blue squares), and z~$>$ 0.5 (solid red stars). The solid, dotted and dashed lines represent the  values, from linear theory, of  ${\cal   R}=R_{\rm  CR}/R_{\rm   bar}= 1.0, 1.2$ and 1.4, respectively. This range corresponds to what it is considered in the literature as fast bars (see Sect.~\ref{introduction}). Bottom panel: outer ring radius  vs. bar semi-major axes radius for both high and low redshift sample, the symbols and lines are the same as the ones represented in the top panel. Notice that all galaxies, regardless of their redshift, clearly fall into the 'fast-bar' region. \label{fig:rbar.rolr}.}
\end{center}
\end{figure*}

\section{Discussion \label{discussion}}

\subsection{Possible caveats}
The  tightness of  the  results shown  in Fig.~\ref{fig:rbar.rolr}  is
somewhat  unexpected considering the  intrinsic uncertainties
inherent to the measurements used  in this work. For instance, we have
assumed that the outer ring  are perfectly circular, which is critical
for the deprojection of the  ring and bar lengths.  In Sect. \ref{olr}
we  justified  the  assumption   of  roundness  for  the  R$_{2}'$  and
R$_{1}$R$_{2}'$ ring  morphologies.  In  addition, we have  taken into
account the  limits of intrinsic axis ratios  given in \cite{buta1995}
to calculate  the errors  in the projected sizes,  and we have demonstrated in  the previous section
that  our   result  does  not   depend  on  the  inclination   of  the
galaxies.  Therefore  we  conclude  that projection  effects  are  not
biasing our results.

Another possible caveat to our result might be that the choice of ring galaxies biases the sample towards a certain pattern speed domain. However, numerical simulations  \citep{byrd1994} have shown that resonant outer rings can be present in both fast and slow bars. All types of ring morphologies are found at different pattern speeds.  In the same work, all types of ring morphologies also developed for different bar strengths. Athanassoula et al. (2010)\nocite{athanassoula2010}, who presented a new theory for ring and spiral formation, argues that there is a connection between the bar strength and the morphology of the rings.  Nevertheless, R$_{2}'$ rings, as are those selected in this work, are located in barred galaxies with similar bar strengths as our galaxies (see Athanassoula et al. 2010). In addition, they show that R$_{2}'$ type rings can be formed in galaxies with fast and slow bars. 

Although the bar size as measured in different rest-frame band passes could be different, it has been recently shown (Gadotti 2011) that in fact the difference in bar size is negligible and therefore we are not introducing a bias by measuring the bars a high-redshift near the $g$-band rest-frame while the low-redshift bar sizes are measured from the SDSS $r-$band.
 

Our low and high redshift samples are by no means complete. Therefore, it is customary to investigate whether this fact is affecting the results presented in Fig. \ref{fig:rbar.rolr}. Since our low and high redshift barred galaxies are similar in size and strength we could be biasing the resulting pattern speeds toward a particular regime. In other words, the fact that we have not observed evolution in the pattern speed could be just  due to the selection of similar fast bars. However, we know from a study of a complete sample of local barred galaxies (Aguerri et al. 2009),  that only 30$\%$ of the local bars show larger lengths than our ringed barred galaxies. Studies of high redshift bars, 0.4~$<$~z~$>$0.8, \citep{jogee2004,barazza2009} have shown that the bar size distribution is similar to that of  local galaxies and; therefore, as discussed before, similar to the bars size range of the bars presented in this work. Therefore, we do not seem to be looking at any special type of bar by analysing ringed galaxies.
As previously mentioned, from numerical models, bars get longer and slower as they age. We can then set a  30$\%$ upper limit to the bars that could have suffered a change in their pattern speed in the last 7 Gyr, assuming that nearby long bars are the end-products of the evolution of fast bars. The remaining 70$\%$ of bars did not  substantially lose angular momentum to the halo, maintaining their pattern speed.  This discussion might be related to an implicit morphological bias, since it remains, even for local galaxies, to derive the pattern speed of bars in very late-type gaseous rich spirals which might suffer an intrinsically different  evolution (e.g., Bournaud and Combes 2002).

\subsection{Comparison with the results from numerical modelling}
 
A recent numerical work \citep{villavargas2010} shows that the evolution of the pattern speed and bar-growth of a bar embedded in a live dark matter halo depends strongly on the gas content. In their simulations, a fixed fraction of the total mass was converted to gas mass, and the evolution of the bar parameters is then followed in time. The presence of gas changes the evolution of both the bar growth and the pattern speed evolution, the addition of gas can stop, or even speed-up, the pattern speed of the bar with time. The bar size is anti-correlated with the disc gas fractions. These gas-rich galaxies would be related to early-type galaxies because the gas leads to larger central mass concentration and therefore larger bulges.  The results we present in this paper  could be in agreement with these gas-rich models. However, the full picture is still unclear since it is observed that longer bars reside in late-type galaxies (e.g., Erwin 2005) which is against the model predictions. Furthermore, we should then explain why all the galaxies should have started with similar gas fractions in their discs. 

There is also the possibility that the bars that we see at $z\sim0.8$ do not survive till the present and therefore, we do not see evolution because the time-scales involved in the formation and destruction are too short.  It has been discussed \cite[see ][]{pfenniger1990, bournaud2002, bournaud2005} that gas-rich bars, i.e., late-type spirals, are short lived, with lifetimes of 1-2 Gyr. This short time scale would mimic a lack of evolution of long-lived bars; however, the galaxies in the sample show morphologies typical of early type spirals and there is evidence, from stellar population studies \citep{perez2009, sanchezblazquez2011}, that bars in early type galaxies are long-lived. If this is the case, and most of the ring galaxies we observe present long-lasting bars,  it would imply that bars cannot have grown in time and kept being in the fast speed regime without increasing significantly in size. Therefore, the fact that we see the ring radius and the bar size covering the same size range at all redshift, and moreover large bars at high-redshift, implies that bars do not grow significantly in size with time. 

The result shown in Fig.~\ref{fig:rbar.rolr} implies that bars have not evolved considerably, neither in size nor in pattern speed, since around the time when the Universe was half its present age.  Most numerical simulations obtain bars that evolve with time, getting longer and stronger while slowing down (e.g., Debattista \&Sellwood 1998; Athanassoula 2003). This effect is mostly due to the angular momentum exchange of the bar-disc system with the dark matter halo.  Thus, the fact that bars are compatible with fast rotators at all redshifts indicates that the angular momentum exchange between the bar and halo has not been important enough in the last 7 Gyr to slow down bars. If the  pattern speed can be used to set constrains to the halo-to-disc mass ratio, these results might imply that  the discs in the high surface-brightness galaxies of our sample are maximal.


\section{Summary and conclusions}
\label{conclusions}
We  have  analysed  44  low  inclination ringed  galaxies  spanning  a
redshift range between  $0 < z < 0.8$ to  study the possible evolution
of the pattern speed in the last 7 Gyrs.

We calculated for each  galaxy a morphological parameter indicative of
the dynamical state  of their bars. In particular,  we derived whether
they are  fast or slow  rotators. We find  that the bar  pattern speed
does not seem to change with redshift and that all bars are compatible
with being fast bars.

If the  bars analysed  are long-lasting, their  size and  bar strength
have not significantly changed in  time. The fact that, independent of
the  redshift, the  bars  are fast  rotators  and their  size has  not
significantly changed  in time could  have also large  implication for
bar evolution  models that mostly predict  a bar growth  with time. It
has  been argued  that  the  exchange  of angular  momentum with  a
centrally  dense halo causes  the bar  to  evolve; however  the
present   results   might   imply   that   the  disc   in   the   high
surface-brightness galaxies is maximal and the central mass density is
dominated  by  the  stellar  component which would  lower  the  angular
momentum exchange between the disc and  the halo and slow down the bar
evolution (e.g., Debattista \&  Sellwood, 2000, but for a different
conclusion, see Athanassoula, 2003).

This  is the  first time  that the  pattern speed  evolution  has been
investigated  from  the  observational  point of  view.   The  results
presented here place strong constrains on the bar evolution models.

\begin{acknowledgements}
We thank the referee for his/her useful comments. We would like to thank Victor Debattista for the careful reading of the manuscript. I.P. was supported by the Spanish Ministry of Science and Innovation (MICINN) (via grants AYA2010-21322-C03-02, AYA2010-21322-C03-03, AYA2007-67625-C02-02 and Consolider-Ingenio CSD2010-00064) and 
by the Junta de Andaluc\'ia (FQM-108). JALA and JMA were supported by the projects AYA2010-21887-C04-04 and by the Consolider-Ingenio 2010 Program grant CSD2006-00070.

      \end{acknowledgements}

\bibliography{patternhighz}
\bibliographystyle{natbib}

\appendix
 
\section{Sample galaxies}
\begin{figure*}
\begin{center}
\resizebox{0.9\textwidth}{!}{\includegraphics[angle=0]{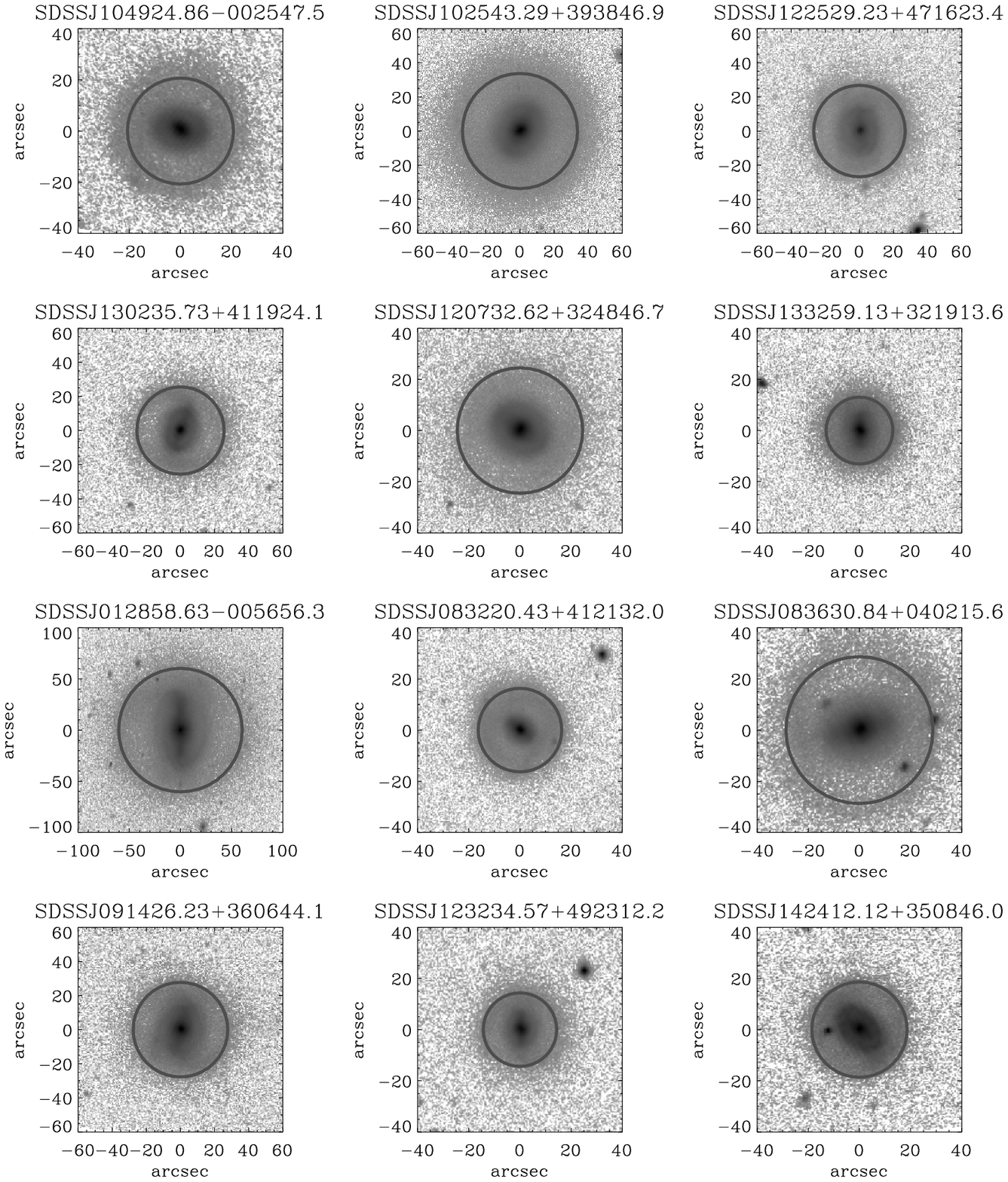}}
\caption{Low redshift sample with black solid circles indicating the ring size. \label{appendix1}}
 \end{center}
\addtocounter{figure}{-1}
\end{figure*}
\begin{figure*}
\begin{center}
\resizebox{0.9\textwidth}{!}{\includegraphics[angle=0]{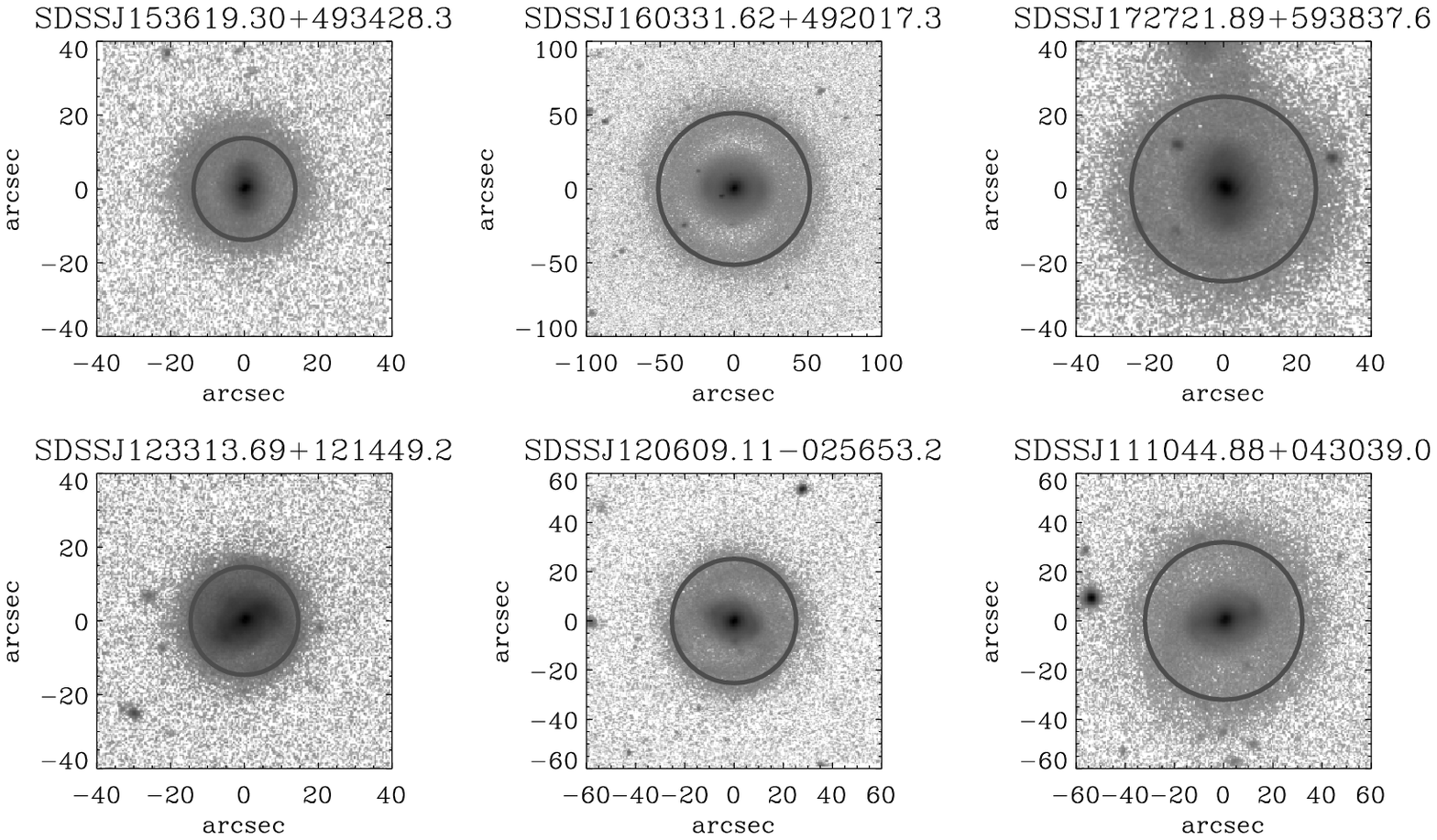}}
\caption{Low redshift sample with black solid circles indicating the ring  size.}
 \end{center}
\end{figure*}
\begin{figure*}
\begin{center}
\resizebox{0.9\textwidth}{!}{\includegraphics[angle=0]{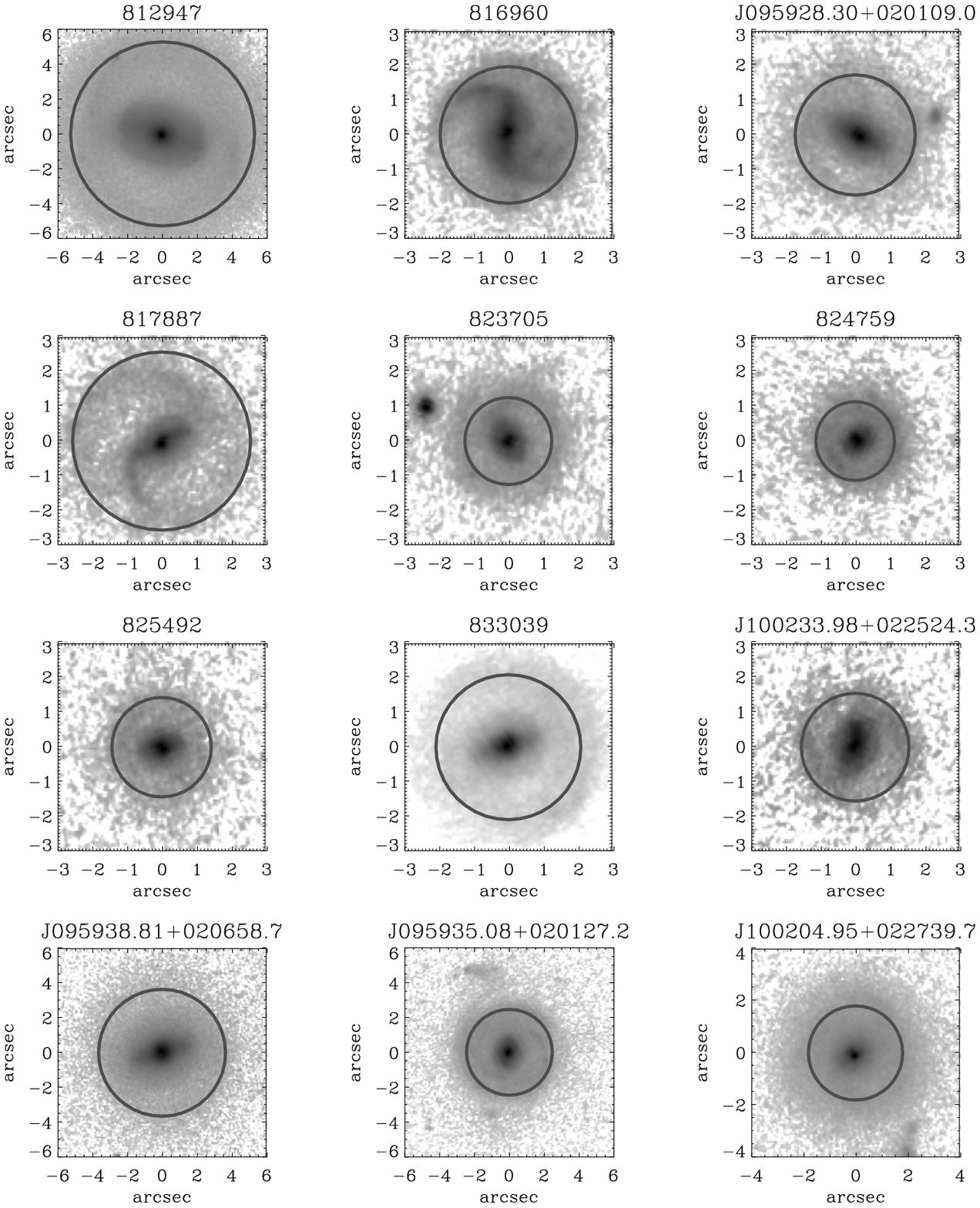}}
\caption{High redshift sample with black solid circles indicating the ring size.\label{appendix2}}
 \end{center}
\end{figure*}
\addtocounter{figure}{-1}
\begin{figure*}
\begin{center}
\resizebox{0.9\textwidth}{!}{\includegraphics[angle=0]{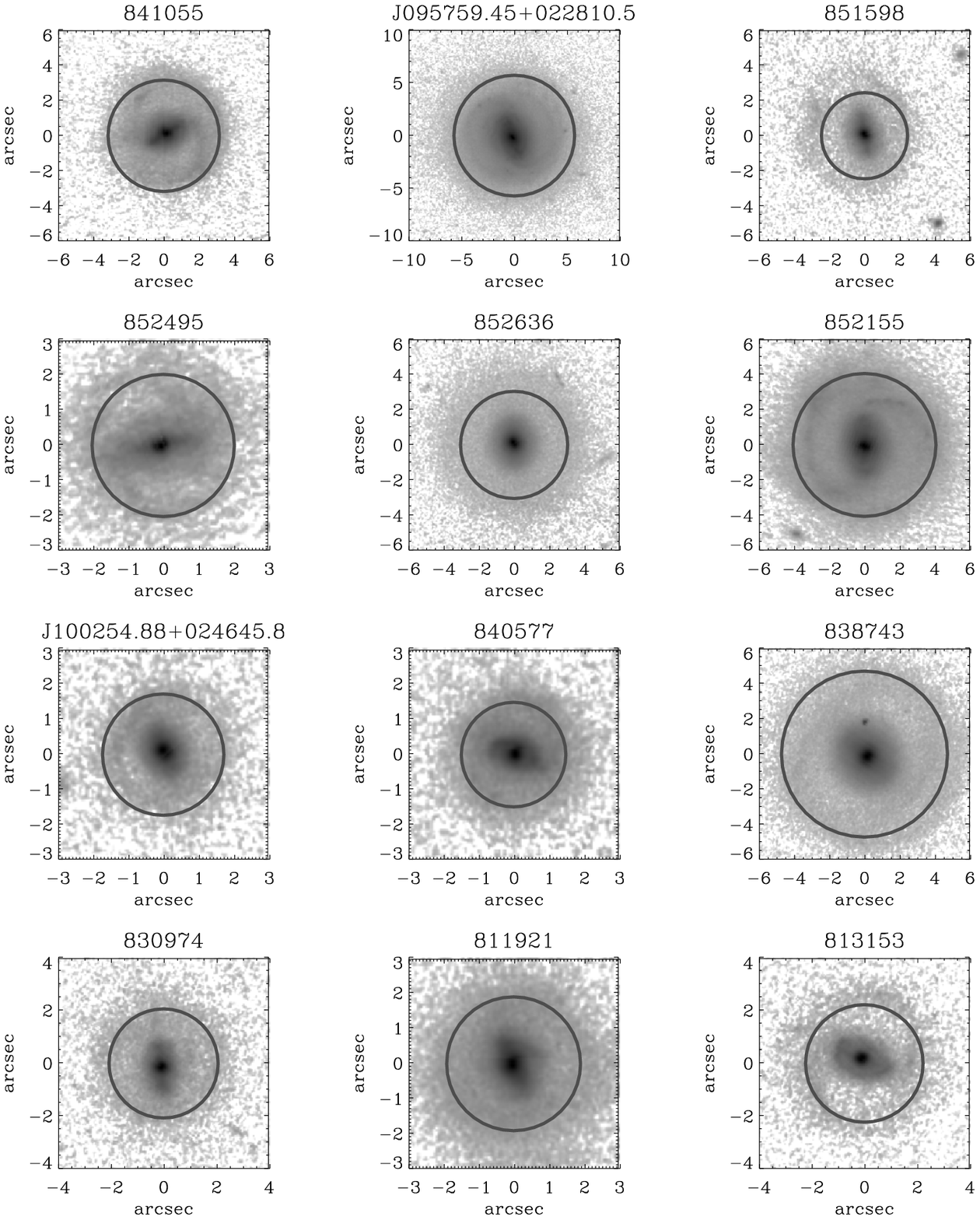}}
\caption{High redshift sample with black solid circles indicating the ring size.}
 \end{center}
\end{figure*}
\addtocounter{figure}{-1}
\begin{figure*}
\begin{center}
\resizebox{0.9\textwidth}{!}{\includegraphics[angle=0]{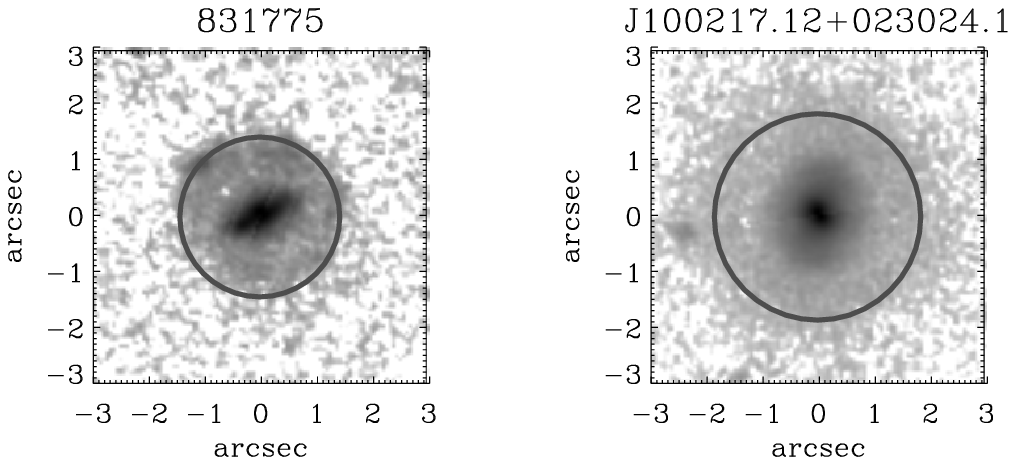}}
\caption{High redshift sample with black solid circles indicating the ring size.}
 \end{center}
\end{figure*}
\end{document}